\documentclass[aps,prl,twocolumn,superscriptaddress,secnumarabic]{revtex4-2}

\usepackage{graphicx}
\usepackage[utf8]{inputenc}
\usepackage{float}
\usepackage{comment}
\usepackage[dvipsnames]{xcolor}
\usepackage{amsmath}

\begin{document}



\title{Spin-orbital mixing in the topological ladder of the two-dimensional metal PtTe$_2$}


\author{M. Qahosh}
\affiliation{Peter Gr{\"u}nberg Institut (PGI-6), Forschungszentrum J{\"u}lich GmbH,
52428 J{\"u}lich, Germany}

\author{M. Masilamani}
\affiliation{New Technologies-Research Center, University of West Bohemia, 30614 Pilsen, Czech Republic}
\affiliation{Experimentelle Physik VII and Würzburg-Dresden Cluster of Excellence ct.qmat, Universität Würzburg, Würzburg, Germany}

\author{H. Boban}
\affiliation{Peter Gr{\"u}nberg Institut (PGI-6), Forschungszentrum J{\"u}lich GmbH,
52428 J{\"u}lich, Germany}

\author{Xiao Hou}
\affiliation{Peter Gr{\"u}nberg Institut (PGI-6), Forschungszentrum J{\"u}lich GmbH,
52428 J{\"u}lich, Germany}

\author{G. Bihlmayer}
\affiliation{Peter Gr{\"u}nberg Institut (PGI-1), Forschungszentrum J{\"u}lich and JARA, 52428 J{\"u}lich, Germany}

\author{Y. Mokrousov}
\affiliation{Peter Gr{\"u}nberg Institut (PGI-1), Forschungszentrum J{\"u}lich and JARA, 52428 J{\"u}lich, Germany}

\author{W. Karain}
\affiliation{Department of Physics, Birzeit University, PO Box 14, Birzeit, Palestine}

\author{J. Min\'ar}
\affiliation{New Technologies-Research Center, University of West Bohemia, 30614 Pilsen, Czech Republic}

\author{F. Reinert}
\affiliation{Experimentelle Physik VII and Würzburg-Dresden Cluster of Excellence ct.qmat, Universität Würzburg, Würzburg, Germany}

\author{J. Schusser}
\affiliation{New Technologies-Research Center, University of West Bohemia, 30614 Pilsen, Czech Republic}
\affiliation{Experimentelle Physik VII and Würzburg-Dresden Cluster of Excellence ct.qmat, Universität Würzburg, Würzburg, Germany}

\author{C. M. Schneider}
\affiliation{Peter Gr{\"u}nberg Institut (PGI-6), Forschungszentrum J{\"u}lich GmbH,
52428 J{\"u}lich, Germany}
\affiliation{Fakult{\"a}t f{\"u}r Physik, Universit{\"a}t Duisburg-Essen, 47048 Duisburg, Germany}
\affiliation{Physics Department, University of California, Davis, CA 95616, USA}

\author{L. Plucinski}
\email{l.plucinski@fz-juelich.de}
\affiliation{Peter Gr{\"u}nberg Institut (PGI-6), Forschungszentrum J{\"u}lich GmbH,
52428 J{\"u}lich, Germany}
\affiliation{Institute for Experimental Physics II B, RWTH Aachen University, 52074 Aachen, Germany}

\date{\today}

\begin{abstract}
We visualize the topological ladder and band inversions in PtTe$_2$ using spin-polarized photoemission spectroscopy augmented by three-dimensional momentum imaging. This approach enables the detection of spin polarization in dispersive bands and provides access to topological properties beyond the reach of conventional methods. Extensive mapping of spin-momentum space reveals distinct topological surface states, including a surface Dirac cone at the binding energy \(E_B \sim 2.3\) eV and additional states at \(E_B \sim 1.6\) eV, \(E_B \sim 1.0\) eV, and near the Fermi level. The electronic structure analysis demonstrates strong hybridization between Pt and Te atomic orbitals, confirming the nontrivial topology of these surface states. Furthermore, by comparison to one-step model photoemission calculations, we identify a robust correlation between the initial-state and measured spin polarizations while revealing asymmetries in specific experimental spin textures. These asymmetries, absent in the initial states due to symmetry constraints, arise from the breaking of time-reversal symmetry during the photoemission process, emphasizing the crucial influence of symmetries on experimental signatures of topology.
\end{abstract}

\maketitle

\section{Introduction}

1T transition-metal dichalcogenides (TMDCs) form a class of materials that exhibit numerous important electronic phenomena, including correlated and topological phases \cite{Clark2018, Li2017, Bahramy2018, Nicholson2021, Ghosh2019, Mukherjee2020}. In particular, Pd and Pt selenides and tellurides exhibit a so-called {\it topological ladder} in the chalcogen-derived $p$-band manifold \cite{Bahramy2018} with several resulting spin-polarized topological states \cite{Clark2018}. The determination of topological properties of these materials requires probing spin-momentum dependence over the full Brillouin zone in order to access the Bloch wave functions, including their spin and orbital characters, as well as phases, that enter the quantum geometric tensor \cite{Gianfrate2020,Kang2024}. Angle-resolved photoemission spectroscopy (ARPES) is the most powerful technique for experimental electronic band structure determination~\cite{Sobota2021}, and can access, albeit indirectly~\cite{Henk2018}, spin band characters through its spin-polarized (SARPES) variant.

Compounds containing heavy elements typically exhibit spin-polarized surface states that exist in the local gaps of the projected band structure and emerge through band inversion at time-reversal invariant momenta (TRIM) points, therefore have the same origin as topological ones \cite{Yan2015}. The topological ladder of highly spin-polarized spin-momentum-locked states in PtTe$_2$ stems from several such band inversions \cite{Bahramy2018,Clark2018}, with additional effects due to hidden spin polarization \cite{Zhang2014,Clark2022} also present. We demonstrate how the interatomic interference effects due to strong orbital intermixing \cite{Heider2023}, augmented with spin-orbit-scattering, lead to complex photoemission spin textures that reflect surface symmetries and carry rich information on amplitude and phases of initial wave functions.
Our findings underscore the utility of SARPES momentum maps for visualizing the topological ladder in TMDCs and exploring the topological properties of quantum materials more broadly.


\begin{figure*}
 \centering
     \includegraphics[width=16cm]{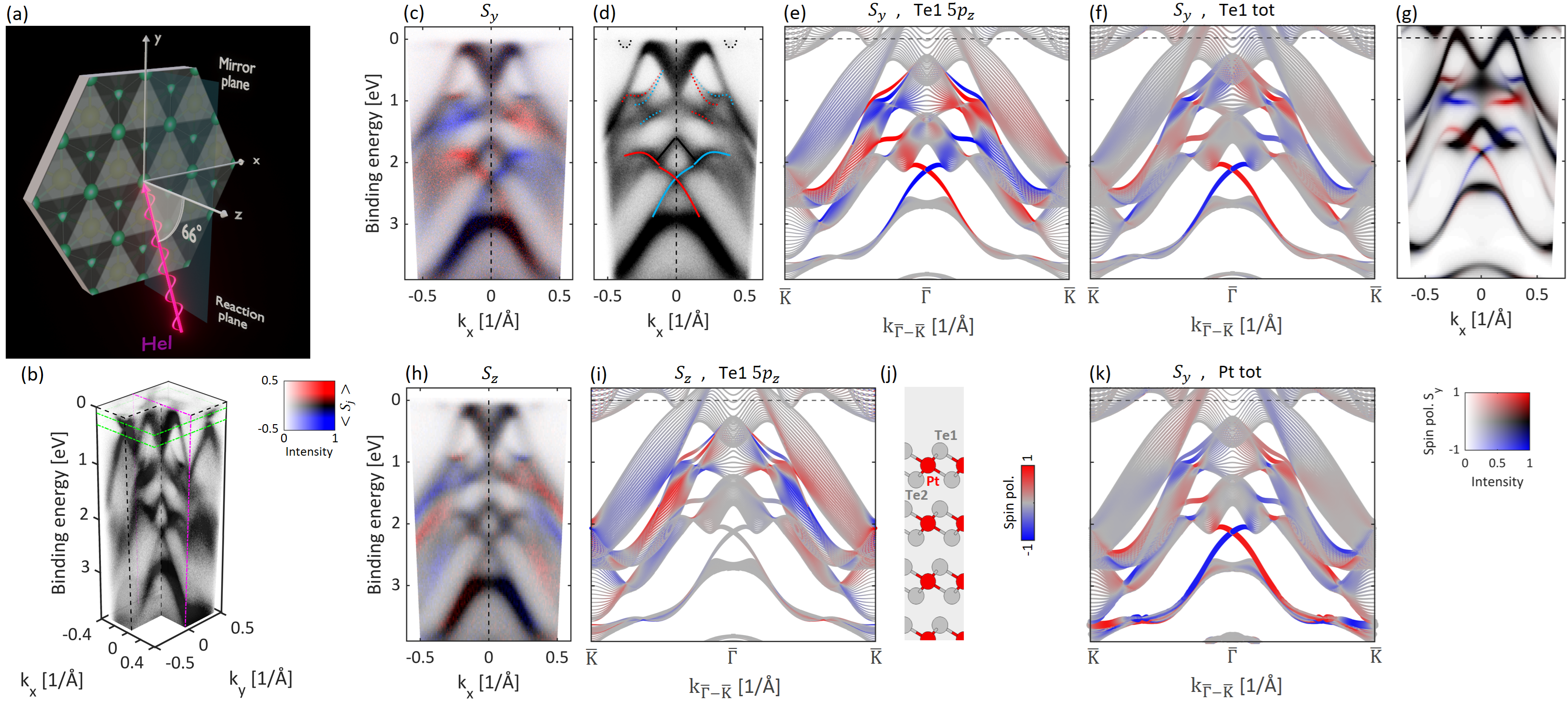}
     \caption{(a) Experimental geometry where the reaction plane coincides with the $\mathcal{M}_x$ mirror plane of the sample's surface. (b) Spin-integrated experimental 3D data. (c) Spin-polarized map for the cut indicated by the magenta frame in (b), color indicates $S_y$. (d) Identification of topological ladder features present in (c) on top of a spin-integrated map. (e) Related {\it ab~initio} spin polarized surface electronic structure calculation, where line thickness indicates the weight and the color the spin expectation value $S_y$ for surface Te $5p_z$ orbitals. (f) Same as (e) but for the total weight on surface Te atom. (g) Simulation of (c) calculated using SPR-KKR \cite{Ebert2011} packet. (h) Same as (c) but for $S_z$.  (i) Same as (e) but for $S_z$. (j) Schematic indication of the surface Te1 atom,  the outermost Pt atom, and the third layer Te2 atom. (k) Same as (f) but for the Pt atom.}
     \label{Fig1:Symmetry}
\end{figure*}

\section{Results}

The surface of 1T-PtTe$_2$ exhibits 3 mirror planes which are oriented along the $\overline \Gamma \overline M$ reciprocal directions. Figure \ref{Fig1:Symmetry} (a)-(g) shows SARPES maps and {\it ab~initio} calculations for the experimental geometry depicted in (a), where one of the sample mirror planes, $\mathcal M_x$, is preserved. Figure \ref{Fig1:Symmetry} (b) shows the experimental spin-integrated band structure, with the magenta frame depicting the plane in which the SARPES map in (c) for the spin expectation value $S_y$ (refer to coordinate system definition in Fig.~\ref{Fig1:Symmetry} (a)) has been measured. Since $S_y$ is parallel to $\mathcal M_x$, through the axial vector mirror reflection rules, the map in Fig.~\ref{Fig1:Symmetry} (c) must obey $S_y(k_x,k_y,E) = -S_y(-k_x,k_y,E)$, which it indeed does. Figure \ref{Fig1:Symmetry} (c) exhibits several spin polarized states related to the topological ladder, which for clarity are depicted by solid lines in (d) on top of the spin-integrated map. Additional dashed lines depict surface resonances and other pronounced features. All of these are reproduced in the {\it ab initio} calculation in Fig.~\ref{Fig1:Symmetry} (e), where the false colormap and the thickness of the lines refers to spin polarization and partial charge of $5p_z$ orbital of the surface Te atom. An additional calculation in Fig.~\ref{Fig1:Symmetry} (f) shows the  total charge on surface Te atoms, where the comparison to the experiment is less favorable than in (e), especially in the region of ladder states at $\approx 1$ eV. Figure \ref{Fig1:Symmetry}~(g) shows the SARPES spectrum related to (c) calculated with the SPR-KKR package \cite{Ebert2011} using a free-electron final state and demonstrating excellent agreement for virtually all experimental features.

Figs.~\ref{Fig1:Symmetry} (h) and (i) show similar experimental and theoretical results as (c) and (e) but for the out-of-plane spin expectation value $S_z$, where again a favorable agreement is found. For convenience, Fig.~\ref{Fig1:Symmetry} (j) schematically depicts the outermost Te and Pt atoms from which the majority of the photoemission signal originates through the method's surface sensitivity, and (k) shows the theoretical $S_y$ calculation for that Pt atom, indicating that the topological ladder wave functions are strongly mixed between the surface-most Pt and Te sites. Taking into account the photoionization cross-sections \cite{YehLindau,Suppl}, photoelectron mean free path, and expected angular distributions from atomic orbitals at the experimental incident angle of $\theta_{h\nu} = 66^\circ$, the dominant photocurrent contributions for the topological ladder states are expected from both Te $5p_z$ and Pt $5d_{z^2}$ states.

The Dirac cone states in Fig.~\ref{Fig1:Symmetry} (c) at $E_B \approx 2.3$ eV are strongly spin polarized, which is consistently reproduced in panels (e), (f), (g), and (k), therefore their $S_y$ polarization is robust. However, the difference between Te $5p_z$ partial charges in Fig.~\ref{Fig1:Symmetry} (e) and total charge calculations in (f) and (k), in particular for the ladder states at $\approx 1$ eV, indicates their complex spin texture in real space.

The in-plane spin texture in PtTe$_2$ follows from a Rashba-like mechanism due to the out-of-plane potential gradient. On the other hand, the out-of-plane spin texture results from the intra-layer in-plane dipoles, which exist along directions orthogonal to the mirror planes \cite{Clark2022}. Despite the different origins of the two effects, the $S_y$ and $S_z$ spin polarizations in Fig.~\ref{Fig1:Symmetry} (c) and (h), respectively, appear to be similar, with the main difference being a small $S_z$ polarization in the Dirac cone, theoretically reproduced in (j).

\begin{figure*}
 \centering
     \includegraphics[width=14cm]{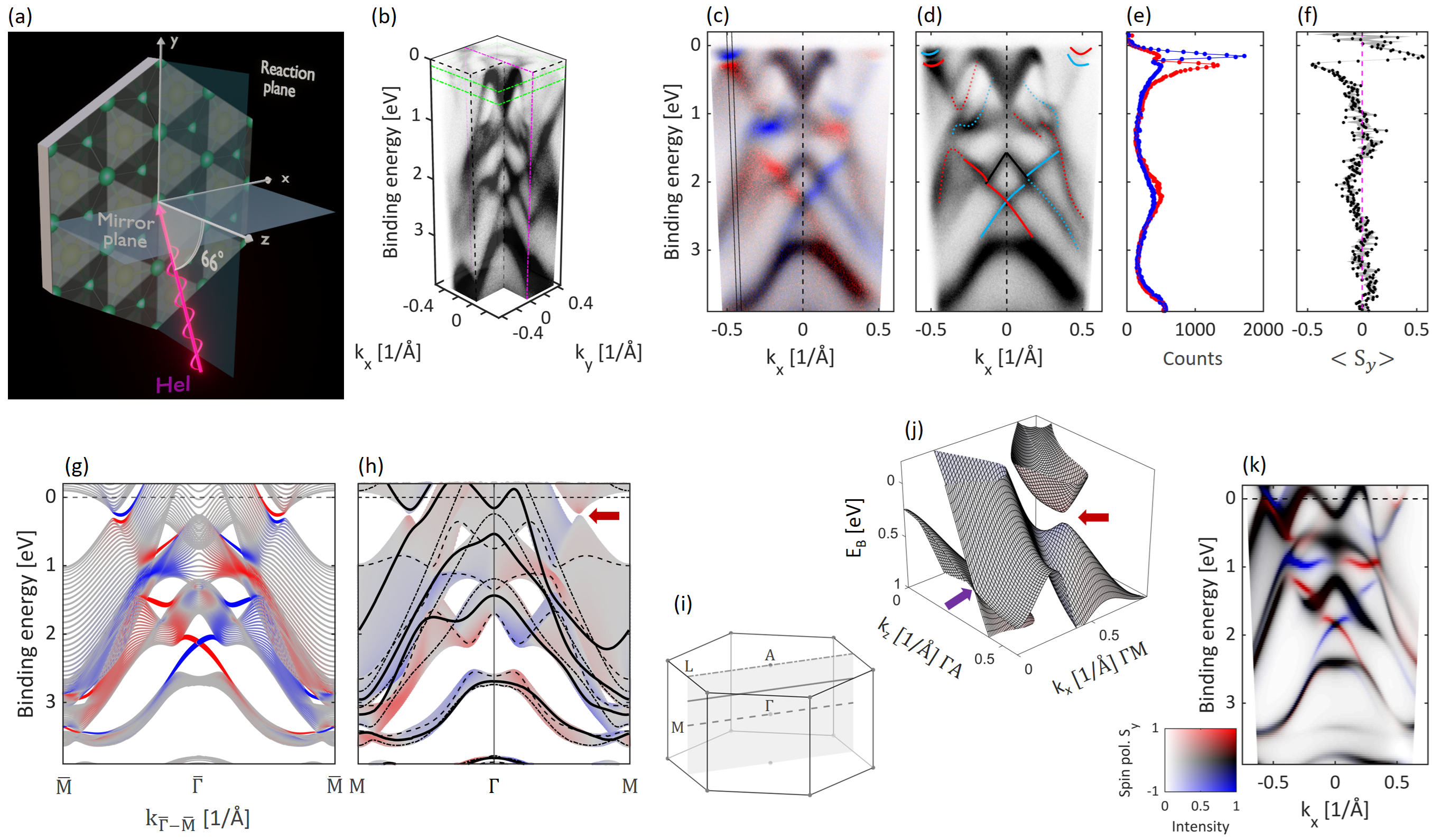}
     \caption{(a) Experimental geometry where the reaction plane is orthogonal to the $\mathcal{M}_y$ mirror plane of the sample's surface. (b) Spin-integrated experimental 3D data. (c) Spin-polarized map for the cut indicated by the magenta frame in (b), color indicates $S_y$. (d) Identification of topological ladder features present in (c) on top of a spin-integrated map. (e) Spin-polarized EDCs for the region in (c) enclosed by the black lines. (f) Related spin polarization. (g) {\it Ab~initio} spin polarized surface electronic structure calculation for where line thickness indicates the weight and the color the spin expectation value $S_y$ for surface Te~$5p_z$ orbitals. (h) Related spin-polarized bulk projected band structure with the dashed line showing bands along $M \Gamma M$, the dashed dotted line along $LAL$, and with the solid line showing bands along a line that cuts through the middle of the $\Gamma A$ line, as shown in (i). (j) Bulk band structure for $E_B(k_x,k_z)$. Dark red arrows in (h) and (j) indicate the local bulk band gap, while the purple arrow in (j) indicates the type-II bulk Dirac cone. (k) One-step photoemission model calculation related to the map (c).}
     \label{Fig2:Asymmetry}
\end{figure*}

Figure~\ref{Fig2:Asymmetry} shows SARPES maps and {\it ab~initio} calculations for the experimental geometry depicted in panel (a), where none of the sample mirror planes is preserved. Figure \ref{Fig2:Asymmetry}~(b) shows experimental spin-integrated band structure, with the magenta frame depicting the plane along which the SARPES map in (c) for the spin expectation value $S_y$ (refer to coordinate system definition in Fig.~\ref{Fig2:Asymmetry}~(a), compared to Fig.~\ref{Fig1:Symmetry} the sample has been rotated) has been measured. The identified topological ladder states are depicted in Fig.~\ref{Fig2:Asymmetry}~(d), while (e)-(f) show the energy distribution curve (EDC) and spin-polarization in the region between the two solid lines in (c) where surface state near the Fermi level with $S_y$ over 50\% are located. States identified in Fig.~\ref{Fig2:Asymmetry} (d) are in good overall agreement with the spin-polarized calculation for surface-most Te $5p_z$ orbital shown in panel (g).

In Fig.~\ref{Fig2:Asymmetry}~(g), the $S_y$ spin polarization is antisymmetric between  $S_y(E_B,k_x,k_y)$ and $S_y(E_B,-k_x,k_y)$. However, since the experimental setup of Fig. \ref{Fig2:Asymmetry}~(a) does not have any mirror planes, no such rules relating $S_y(E_B,k_x,k_y)$ with $S_y(E_B,-k_x,k_y)$ exist for the experimental SARPES map, and the maps in (c)-(d) exhibit clear asymmetries, both regarding the spin polarization and the energy position of the dispersive features. For surface states this phenomenon, previously revealed for WTe$_2$ \cite{Heider2023}, is related to the interatomic interference and time-reversal symmetry breaking in the photoemission process, while for bulk states to the asymmetric initial band structure dispersions for generic values of $k_z$ momenta, as shown in Fig.~\ref{Fig2:Asymmetry} (h)-(i), through approximate $k_z$ selectivity of ARPES \cite{Strocov2023}. Importantly, wave function mixing between different sites is necessary to induce the effect for surface states, which provides an additional confirmation for such mixing in PtTe$_2$ topological ladder. Figure \ref{Fig2:Asymmetry} (j) shows calculated $E_B(k_x,k_z)$ bulk bands, highlighting the local band gap where the strongly spin-polarized state visualized in (e) is located, as well as the type-II bulk Dirac cone \cite{Yan2017}. Figure \ref{Fig2:Asymmetry} (k) presents the SPR-KKR \cite{Ebert2011} calculation corresponding to the map in (c), incorporating the aforementioned effects and confirming the strong $\pm k_x$ asymmetries observed in the experiment.

\begin{figure*}
 \centering
     \includegraphics[width=14cm]{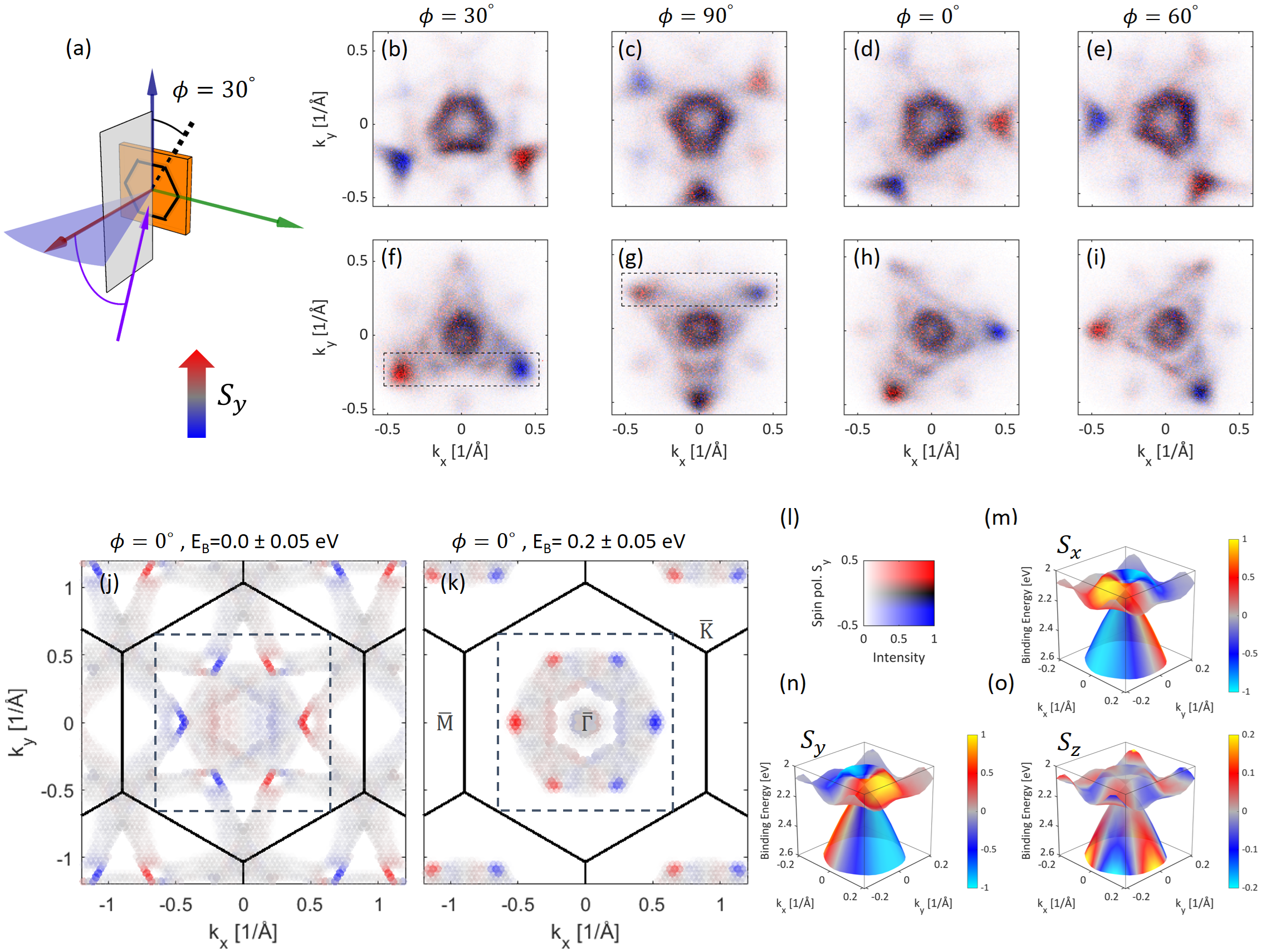}
     \caption{(a) Experimental geometry indicating the rotation of the sample with respect to the $z$-axis by an angle $\phi$. (b)-(e) Experimental SARPES maps for $\phi = 30^\circ, 90^\circ, 0^\circ$, and $60^\circ$, respectively, at the binding energy $E_B = 0.1$ eV. (f)-(i) Same as (b)-(e) but at $E_B = 0.3$ eV. (j)-(k) Theoretical spin-polarized constant energy maps at $E_B = \pm 0.05$ eV and $E_B = 0.2\pm 0.05$ eV, respectively. (l) Colormap used in (b)-(k), the spin-quantization axis is for $S_y$. (m)-(o) Calculated dispersions of the Dirac cone, with colors indicating $S_x$, $S_y$, and $S_z$ quantization axes, respectively, with false color scales indicated. Note that the color scale has been saturated to 20\% in (o).}
     \label{fig3:kxky}
\end{figure*}

Figure \ref{fig3:kxky}~(a)-(i) shows the experimental geometry and SARPES constant energy maps (CEMs) for two different binding energies (as indicated by green contours in Fig.~\ref{Fig1:Symmetry}~(b) and Fig.~\ref{Fig2:Asymmetry}~(b)) and different sample rotations $\phi$, as defined in panel (a). The related calculated initial band structure maps are shown in Fig.~\ref{fig3:kxky}~(j) and (k). These maps aim at imaging strongly spin polarized states near the Fermi level, see e.g.  Fig.~\ref{Fig2:Asymmetry}~(e). Indeed, in all maps one can see spin-polarization sign inversion between $E_B = 0.1$ and $0.3$ eV, however, strong effects related to incident light direction are observed. Due to trigonal symmetry of PtTe$_2$, when $\phi$ is changed by $60^\circ$ the SARPES results are different. This makes the maps of Fig.~\ref{fig3:kxky}~(b) and (c), where the mirror plane coincides with the reaction plane, not only appearing rotated by $\Delta \phi = 60^\circ$ but also exhibiting different intensity pattern and different strength of spin polarization (same for (f) and (g)). For the geometry where the mirror plane is orthogonal to the reaction plane, the maps of Fig. \ref{fig3:kxky} (d) and (e) are expected to be related to each other by the $\mathcal M_x$ transformation, which is indeed observed in experiment (same for (h) and (i)).

Figure \ref{fig3:kxky} (m)-(o) show the {\it ab initio} band structure depicting the spin texture of the surface Dirac cone. One can see that besides the strongly spin polarized in-plane Rashba-like spin-momentum locked spin texture, an out-of-plane spin polarization of $\approx 20\%$ is also predicted in the $S_z$ component, similar to the polarization in the warped Dirac cone of topological insulators \cite{Eremeev2012}.

\begin{figure}
 \centering
     \includegraphics[width=6cm]{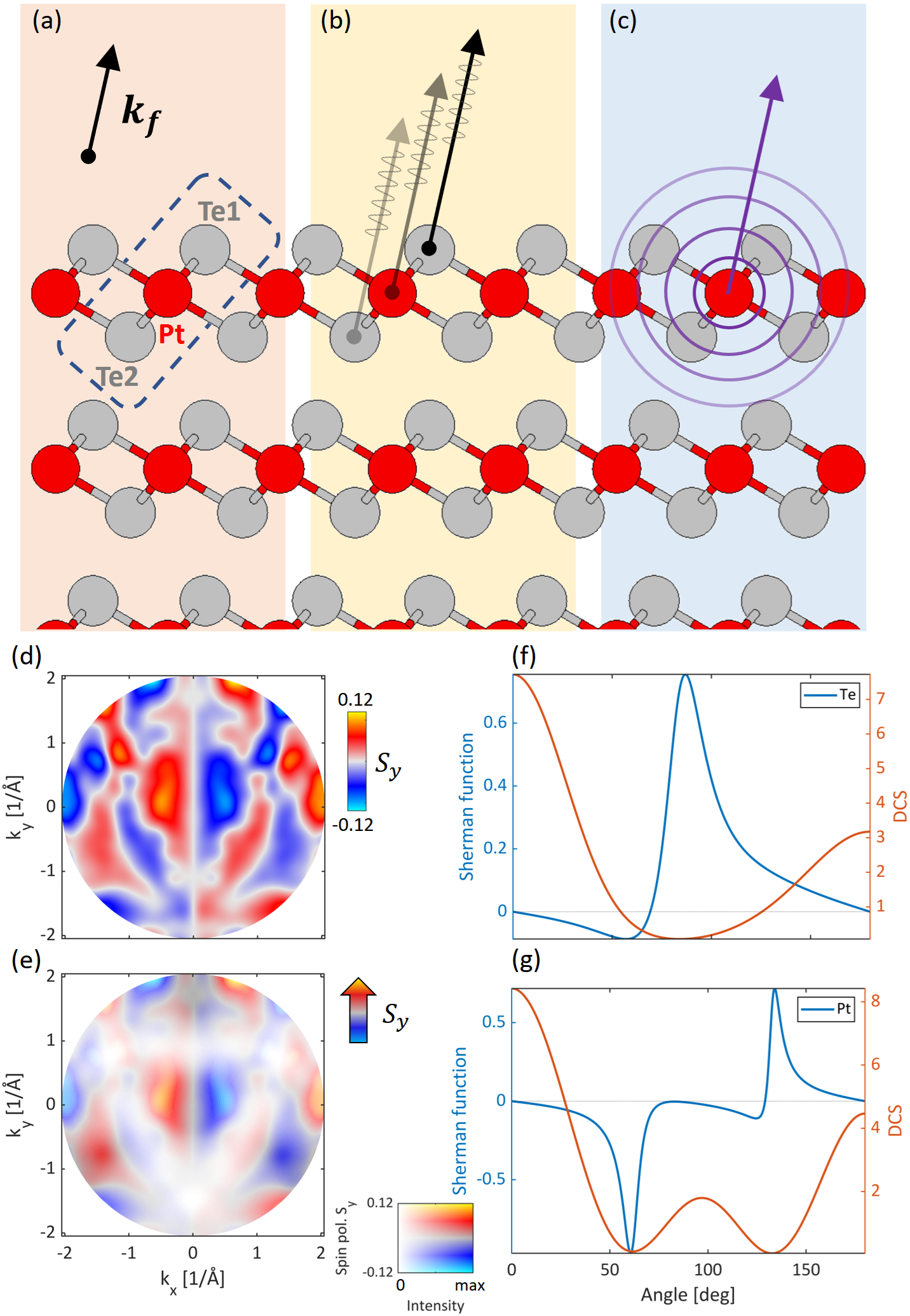}
     \caption{(a) Indication of the 3 atoms in the unit cell of the outermost PtTe$_2$ layer. (b) The photoemission signal can be considered as a far-field coherent sum of electron waves originating from different atomic sites. (c) Spherical wave originating from the Pt site. (d) $S_y$ spin-polarization in the photoelectron diffraction pattern for SOC-scattering of the $s$-wave emission from the outermost Pt site in bulk PtTe$_2$ at $E_{kin} = 16$ eV. (e) Same as (d), but with the special 2D colormap that simultaneously represents both spin polarization and intensity. (f),(g) Differential cross-sections (DCS) and Sherman functions corresponding to the muffin-tin potentials employed in the photoelectron diffraction (PED) calculation, for Te and Pt atoms respectively.}
     \label{Fig4:EDAC}
\end{figure}

Figure \ref{Fig4:EDAC} provides a detailed breakdown of various components appearing in the SARPES process. As depicted in Fig.~\ref{Fig4:EDAC} (a), due to the short inelastic mean free path, in photoemission one can consider three outermost layers \cite{Beaulieu2020}, and due to the in-plane periodicity, it is sufficient to consider only the three atoms in the unit cell (of course further layers must be considered in accurate modeling). Figure \ref{Fig4:EDAC} (b) depicts the interatomic interference process \cite{Heider2023,Krueger2018}, which is sufficient to qualitatively explain the asymmetries of Fig.~\ref{Fig2:Asymmetry} (c). However, even in a tight binding picture, in quantitative simulation, one needs to consider a true spherical wave originating from every participating orbital, which means that multiple scattering, which includes spin-orbit coupling (SOC) scattering, needs to be considered. Pictorially this is indicated in Fig.~\ref{Fig4:EDAC} (c), where a spherical wave, originating from one site, will be scattered by all the sites. Fig. \ref{Fig4:EDAC} (d)-(g) provides insight into how much SOC scattering alone can influence SAPRES maps from PtTe$_2$. In order to estimate spin-polarization due to SOC-scattering, we have performed real-space atomic cluster calculation \cite{Abajo2001}, where we have emitted an unpolarized electron $s$-wave from the Pt atom, as depicted in (c). Since PtTe$_2$ in non-magnetic (i.e. no exchange scattering), without SOC-scattering such wave would not lead to spin polarized electrons, even if multiply scattered. However, with spin-orbit scattering, a spin polarization of up to 15\% is present at certain emission angles. This value is likely smaller in actual SARPES maps, since the angular regions of high spin polarization must coincide with emission angles where bands appear, and, furthermore, coherent sum of emissions from different sites must be considered.

\section{Discussion}

We demonstrate that a qualitative understanding of experimental spin textures of the topological ladder states in PtTe$_2$ requires considering orbital characters that are preferentially excited. Strong asymmetries observed in Fig. \ref{Fig2:Asymmetry} (c) show that the details of measured polarization results from interdependence of several effects that include interatomic interference, perpendicular momentum sensitivity of SARPES, as well as the SOC scattering processes. Some features due to these effects can be modeled using {\it ab initio} methods, as shown in Fig.~\ref{Fig2:Asymmetry} (h) and (j) and in Fig.~\ref{Fig4:EDAC} (d)-(e). Importantly, a favorable agreement to SPR-KKR calculations with the free-electron final state model is obtained, which confirms that in PtTe$_2$ SOC scattering is not critical. Figure \ref{fig3:kxky} demonstrates the crucial importance of considering details of experimental geometry and necessity of measuring SARPES maps over large momentum regions.

Assuming the tight-binding initial state wave function $\psi_i = \sum_j C_j \cdot \phi_j$, where $j \in \{\mathbf{r}_j,n,l,m,s\}$, $\mathbf{r}_j$ are the positions of ions, and $n, l, m,s$ are atomic quantum numbers, the photoemission matrix element can be expressed as $\sum_j C_j \langle \psi_f | \boldsymbol{\varepsilon} \cdot \mathbf{p} | \phi_j \rangle$ \cite{Schattke2003,Krueger2018}. This means that wave function amplitudes and phases encoded in $C_j$ directly modulate  the photoemission signal, which therefore encodes them. Importantly, within this picture, knowledge of $C_j$ allows the determination of quantum geometric tensor \cite{Gianfrate2020,Kang2024}. Nevertheless, care need to be taken in using either a free-electron final state or the independent atomic center approximation in the description of valence band photoemission \cite{Krueger2018,Schusser2022,Schusser2024}.


In summary, we have demonstrated how the properties of topological ladder states are manifested in SARPES spectra. By addressing the challenges posed by intersite orbital mixing and SOC scattering, our work provides a foundation for the quantitative derivation of the quantum geometric tensor in solids using ARPES spectra.

\section{Acknowledgements}

L. P. would like to thank Frank Freimuth for fruitful discussions. M. Q. was supported by the Federal Ministry of Education and Research of Germany in the framework of the Palestinian-German Science Bridge (BMBF grant number 01DH16027). H.B. was supported by the DFG via the Project PL 712/5-1. X. H. was supported by Deutsche Forschungsgemeinschaft (DFG, German Research Foundation) under Germany’s Excellence Strategy - Cluster of Excellence Matter and Light for Quantum Computing (ML4Q) EXC 2004/1 - 390534769. Y.M. acknowledges support  by the EIC Pathfinder OPEN grant 101129641 ``OBELIX'' and by the DFG $-$ TRR 288 $-$ 422213477 (project B06). J.S. and J.M. would like to thank the QM4ST project with Reg. No. CZ.02.01.01/00/22\_008/0004572, cofounded by the ERDF as part of the MŠMT.


%

\end{document}